\documentstyle[prl,aps,multicol,epsfig,amsmath]{revtex}

\begin{document}
\normalsize
\draft
\widetext

\title{Nonthermal fragmentation of C$_{60}$}
\author{H. O. Jeschke}
\address{Institut f{\"u}r Theoretische Physik der Freien
Universit{\"a}t Berlin,
Arnimallee 14, 14195 Berlin, Germany,}
\author{M. E. Garcia and J. A. Alonso}
\address{Departamento de F\'{\i}sica, Universidad de Valladolid, 
47011 Valladolid, Spain.}
\date{\today}
\maketitle

\begin{abstract}
A theoretical study of the subpicosecond fragmentation of C$_{60}$
clusters in response to ultrafast laser pulses is presented. We
simulate the laser excitation and the consequent nonequilibrium
relaxation dynamics of the electronic and nuclear degrees of
freedom. The first stages of the nonequilibrium dynamics are dominated
by a coherent breathing mode followed by the cold ejection of single C
atoms, in contrast to the dimer emission which characterizes the
thermal relaxation. We also determine the nonequilibrium damage
thresholds as a function of the pulse duration. 
\end{abstract}

\begin{multicols}{2}

%\section{Introduction.}

During the last decade much attention has been paid to the study of
the ultrafast fragmentation of molecules and clusters upon excitation
with femtosecond laser pulses\cite{manzwoeste,apbschwerp}. The
sub-picosecond relaxation dynamics of the nuclear degrees of freedom,
as a typical ultrafast phenomenon, becomes interesting when it
presents new features, qualitatively different from the well known
thermal relaxation processes. There are many recent examples of such
new phenomena in solids, like the cold melting of graphite and
silicon\cite{soko:95,downer:92}, ultrafast desorption\cite{wolf:99} or
femtosecond graphitization of diamond\cite{jeschke:99}. For clusters,
the search for ultrafast phenomena presenting nonthermal features is
difficult. The reason for that is the lack of well established thermal
effects from which ultrafast nonthermal phenomena could be
distinguished. There is, however, a very robust thermal phenomenon in
cluster physics: the cooling of vibrationally excited C$_{60}$
clusters through emission of carbon
dimers\cite{smalley:87,scuseria:94}. This remarkable process is
closely related to the cage structure and the high stability of
fullerenes. On large time-scales, successive and stochastic dimer
emission dominates the decay of a hot fullerene.  Now, a fundamental
question regarding the short-time behavior of fullerenes is whether
the dimer-emission is also present in the nonthermal relaxation of
C$_{60}$ after an ultrafast excitation, or if other mechanisms become
dominant. In this letter we show that the dimer emission is not
present in the nonthermal response of C$_{60}$ immediately after
excitation by an intense femtosecond pulse. Instead, the energy
absorbed from the laser pulse concentrates first on a breathing mode
and part of it is then rapidly transferred to single C atoms, which
leave the cluster before any kind of thermalization of the atomic
degrees of freedom can take place. We will refer to this process as
``cold'' fragment emission.

Recently, several experiments on femtosecond excitation of fullerenes
have been
performed\cite{hertel1,hertel2,campbell,hunsche}. Experimental results
suggest that, as expected, the long-time relaxation is dominated by
dimer emission. However, no clear conclusions can be drawn about the
fragmentation behavior at very short times.  Since such illumination
by intense femtosecond pulses involves absorption of many photons by
the cluster, many different electronic excitations are created, which
lead to different relaxation processes and relaxation products.  As a
consequence experimental results yield a superposition of thermal and
nonthermal fragmentation, (multiple) ionization effects and even
Coulomb explosion. Due to the complexity of the problem an exact
treatment of the excitation and relaxation processes is not
possible. Therefore we simplify the problem and focus in this paper on
a particular relaxation mechanism, which results from excitation of
many electron-hole pairs on the neutral cluster.  This means that,
from all possible excitations upon ultrafast absorption of many
photons we only consider those involving excited electrons below the
continuum states. Thus, we do not consider single or multiple
ionization processes, which are, of course, also present.  However,
from phase-space arguments the processes we take into account in this
work should be the most probable ones for not very high laser
intensities.

%\section{Theory.}

In order to describe the nonthermal dynamics of C$_{60}$ we write down
a classical Lagrangian for the atoms, which contains the effect of the
electronic system as a many-body potential $\Phi(\{r_{ij}\},t)$:
\begin{equation}\label{eq:lagrange1}
{\cal L} = \sum_{i=1}^N \frac{m_i}{2} \dot{{\bf r}}_i^2 
- \Phi(\{r_{ij}\},t)\,.
\end{equation}
Here, a number of $N$ atoms with masses $m_i$ at positions ${\bf r}_i$
interact through the potential $\Phi(\{r_{ij}\},t)$, which depends on
the distances $r_{ij} = |{\bf r}_i - {\bf r}_j|$ between the
atoms. Thus, the first term of ${\cal L}$ is the kinetic energy of the
$N$ particles, calculated from the velocities $\dot{{\bf
r}}_i$. $\Phi(\{r_{ij}\},t) = \Phi(H_{\rm el}(\{r_{ij}\}),t)$ is a
complicated functional of the electronic Hamiltonian $H_{\rm el}$. The
equations of motion for a cluster can immediately be derived through
the Euler-Lagrange equations. In the case of the tight-binding (TB)
Hamiltonian employed in this work the gradients of the interaction
potential $\Phi(\{r_{ij}\},t)$ are not available in a closed form and
the equations of motion cannot be integrated analytically. Thus, a
numerical procedure has to be used. We employ the Verlet algorithm in
its velocity form~\cite{verlet:67,haile:92}

For the determination of the interaction potential
$\Phi(\{r_{ij}\},t)$ between the atoms we employ an Hamiltonian $H$
that consists of a tight-binding (TB) part $H_{\rm TB}$ for the
electronic system, the other part being a repulsive potential
$\phi(r_{ij})$ that takes care of the repulsion between the ionic
cores:
\begin{align}\label{eq:hamop1}
  H = &H_{\rm TB} + \sum_{i<j} \phi (r_{ij})\intertext{with}\label{eq:hamop2}
&H_{\rm TB} = \sum_{i\eta} \epsilon_{i\eta} n_{i\eta} + 
\sum_{\substack{ij\eta\vartheta \\ j\not=i}}
t_{ij}^{\eta \vartheta} c_{i\eta}^+ c_{j\vartheta}^{ }\,.
\end{align}
Here, $n_{i\eta}$ represents the occupation number operator for the
orbital $\eta$ of atom $i$, $c_{i\eta}^+$ and $c_{j\vartheta}^{ }$ are fermion
creation and annihilation operators, and the hopping matrix element
has been abridged by $t_{ij}^{\eta \vartheta}$. For the description of carbon,
the $2s$, $2p_x$, $2p_y$ and $2p_z$ orbitals are taken into
account. The angular dependence of the hopping matrix element is
treated following the work of Slater and Koster\cite{slater:54}, while
for the radial part of $t_{ij}^{\eta \vartheta}$ and for the distance dependence
of the repulsive potential $\phi(r_{ij})$ we employ the form proposed by
Xu {\it et al.}\cite{ho:92}. Diagonalization of the Hamiltonian of
Eq.~(\ref{eq:hamop2}) yields the energy spectrum $\{\epsilon_m(\{r_{ij}(t)\})\}$
of the material and thus allows for the calculation of the potential
energy
\begin{equation}\label{eq:enpot}
  \Phi(\{r_{ij}(t)\},t) = \sum_{m} n(\epsilon_m,t) \epsilon_m + 
 \sum_{i<j} \phi (r_{ij}) \,.
\end{equation}
Here, $n(\epsilon_m,t)$ is a time-dependent distribution of the electrons
over the energy levels $\epsilon_m$. Initially, it is given by a Fermi-Dirac
distribution $n^0(\epsilon_m) = 2/(1+\exp{\{(\epsilon_m-\mu)/k_{\rm B}T_{\rm
e}\}})$ at a given electronic temperature $T_{\rm e}$. Its time
dependency that is caused by the absorption of an ultrafast laser
pulse and by subsequent thermalization is calculated according to
\begin{equation}\begin{split}\label{eq:absorpthermal}
\frac{dn(\epsilon_m,t)}{dt} =& \int_{-\infty}^{\infty} d\omega\; 
g(\omega,t-\Delta t) 
\biggl\{\left[ n(\epsilon_m- \hbar \omega, t-\Delta t)  \right.  \\ 
& \left.  + n(\epsilon_m+ \hbar \omega, t-\Delta t)  - 2n(\epsilon_m, 
t-\Delta t)\right] \biggr\}  \\  
 & - \frac{n(\epsilon_m,t) - n^0(\epsilon_m)}{\tau_1} \,.
\end{split}\end{equation}
Here, the laser pulse is characterized by an intensity function
$g(\omega,t)$ which describes the distribution of intensity over time and
energies. Thus, the electronic distribution $n(\epsilon_m,t)$ is at each time
step folded with the current laser intensity function $g(\omega,t)$. This
means that at each time step, the occupation of an energy level $\epsilon_m$
changes in proportion to the occupation difference with respect to
levels at $\epsilon_m - \hbar \omega$ and at $\epsilon_m + \hbar \omega$. We model the complex
processes of electron-electron collisions, that lead to an
equilibration of the electronic system, by a rate equation of the
Boltzmann type for the distribution $n(\epsilon_m,t)$. Thus, with a time
constant $\tau_1$, the distribution $n(\epsilon_m,t)$ approaches a Fermi-Dirac
distribution $n^0(\epsilon_m)$ at a high electronic temperature $T_{\rm
e}$. As we are not aware of a measured relaxation time in C$_{60}$, we
use $\tau_1=10$~fs, a value that was reported for GaAs~\cite{knox:88}.

Now we can determine the forces which are needed for the solution of
the equations of motion by calculating the gradient of the
time-dependent potential $\Phi(\{r_{ij}(t)\},t)$ of Eq.~\ref{eq:enpot}:
\begin{equation}\label{eq:forces}\begin{split}
  {\bf f}_k(\{r_{ij}(t)\},t) = &- \sum_{m} n(\epsilon_m,t) 
\langle m|\nabla_k\: H_{\rm TB}(\{r_{ij}(t)\}) |m \rangle 
	\\&- \sum_{i<j} \nabla_k\:\phi (r_{ij}) \,,
\end{split}\end{equation}
where $\nabla_k \equiv \partial/\partial{\bf r}_k$ and $|m \rangle$ is the eigenvector of $H$
corresponding to eigenvalue $\epsilon_m$. In this equation, a term which
follows from the gradient of the occupations $n(\epsilon_m,t)$ has been
neglected. In Eq.~(\ref{eq:forces}) the Hellman-Feynman theorem has
been used. It is important to keep in mind that we are actually using
a generalization of the adiabatic principle when we consider TB energy
levels with time-dependent fractional occupation numbers
$n(\epsilon_m,t)$. Note that a calculation of the true nonadiabatic evolution of
electronic wave functions is at present only possible for two or three
degrees of freedom, while this work studies the time evolution of $3 N
= 180$ degrees of freedom in the case of C$_{60}$.

%\section{Results.}

Now we present the results for the fragmentation of C$_{60}$ clusters,
which show important differences between thermal and nonthermal
response. As experiments usually provide only electron-emission
spectra or mass spectra of ionized fragments on long time scales, this
theoretical investigation is complementary to the experimental results
in the sense that it can clarify mechanisms and time evolution of the
damage in the clusters during the first stages of the relaxation
process.

\begin{figure}
\includegraphics[width=0.45\textwidth,bb=90 100 520 740]
{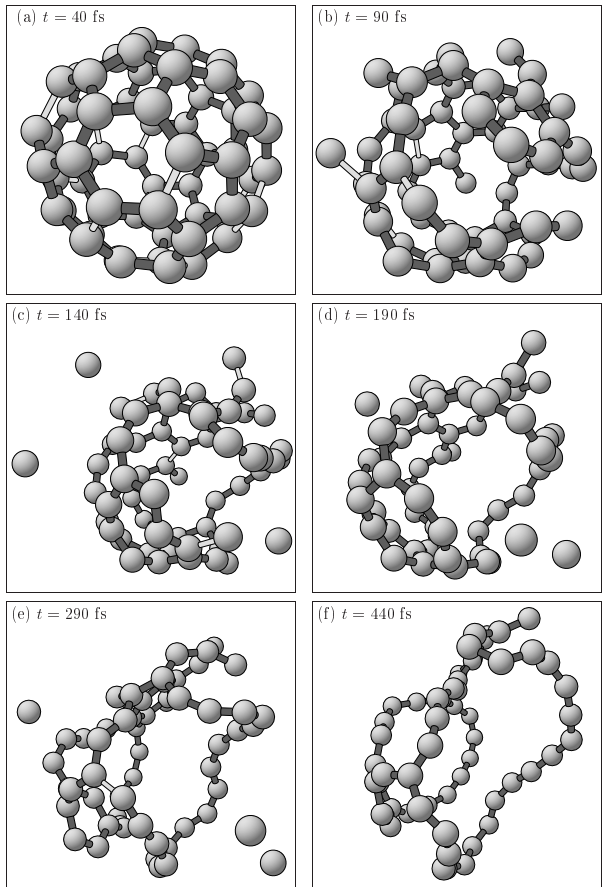}
\caption{
Nonthermal fragmentation of a C$_{60}$ in response to a $\tau=80$~fs
laser pulse. An energy of $E_0 = 3.5$~eV/atom was absorbed. While the
cluster is still intact at the time $t=40$~fs (measured with respect
to the pulse maximum), at $t=90$~fs the cage structure has already
been partially destroyed. In (c) we see three monomers being ejected
from the cluster. A coil of linear carbon chains remains at $t=440$~fs
after the emission of a total of 6 carbon monomers.}
\label{fig:80fs}\end{figure}

A typical fragmentation process is shown in the structure snapshots of
Fig.~\ref{fig:80fs}, which corresponds to an absorbed energy $E_0=
3.5$~eV/atom from a $\tau = 80$~fs laser pulse. While 40~fs after the
pulse maximum, the cluster is still intact, already 50~fs later, the
structure has been torn open and we can see carbon atoms and chains
dangling from the remainders of the C$_{60}$ cage. Again 50~fs later
the emission of three carbon monomers is observed. This emission of
monomers is the dominant initial fragmentation mechanism we obtain for
femtosecond laser pulses. In the further subpicosecond dynamics of the
main fragment four more monomers are emitted. Carbon atoms that have
moved far away from the remaining cluster are not shown in the
subsequent panels. At $t = 440$~fs, a coil of carbon chains has
formed. They stabilize to form three independent chains of 15, 16, and
22 atoms. This fragmentation product is similar to the linked chain
structure found as a result of thermal bond
breaking\cite{tomanek:94}. The C$_{16}$ cluster has a closed ring
structure.  Note that, apart from the process shown in
Fig.~\ref{fig:80fs}, other relaxation mechanisms are possible. For
instance, an absorbed energy of 3.5~eV/atom could be enough to ionize
the cluster several times. We stress again that we only consider
excitations below the continuum. In the experiments, however, both
kind of mechanisms are present.

\begin{figure}
\includegraphics[width=0.45\textwidth,bb=90 100 520 740]
{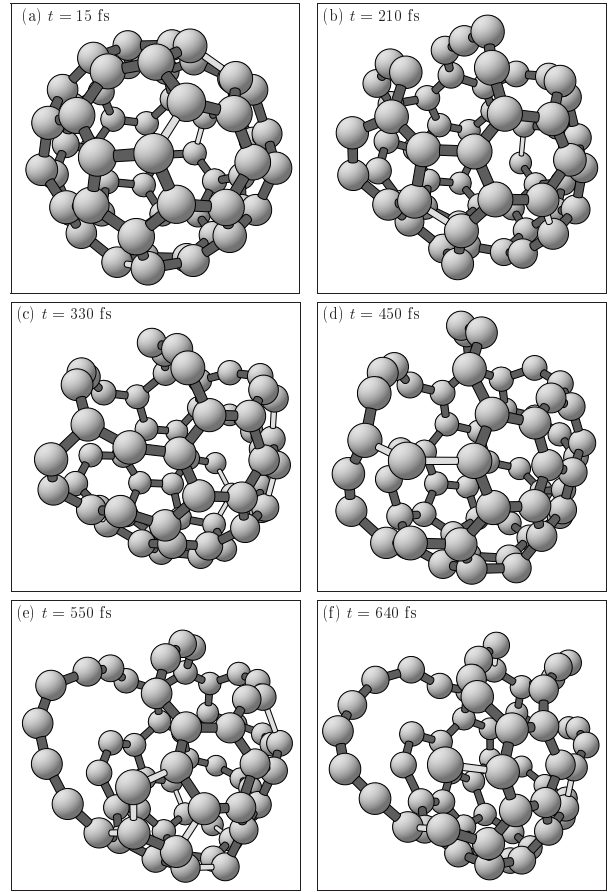}
\caption{
Nonthermal isomerization of a C$_{60}$ in response to a $\tau=5$~fs
laser pulse. An energy of $E_0 = 2.3$~eV/atom was absorbed. Numerous
bonds are broken already at the time $t=210$~fs after the pulse. The
damage develops into a long carbon chain attached on both sides to the
remains of the C$_{60}$ cage structure.}
\label{fig:5fs}\end{figure}

In Fig.~\ref{fig:5fs}, snapshots of a C$_{60}$ cluster isomerization
as a consequence of the absorption of $E_0 = 2.3$~eV/atom from a
$\tau=5$~fs laser pulse are shown. This value of the absorbed energy is
close to the damage threshold of $t_d = 2.1$~eV/atom and thus the
isomerization process takes place on a fairly long time scale of a few
hundred fs. At $t = 200$~fs after the laser pulse maximum the first
breaking of bonds is taking place. This damage to the cluster develops
over the following 400~fs into a chain of carbon atoms that is
attached at both ends to the original molecule.

\begin{figure}
\includegraphics[height=0.45\textwidth,angle=-90]
{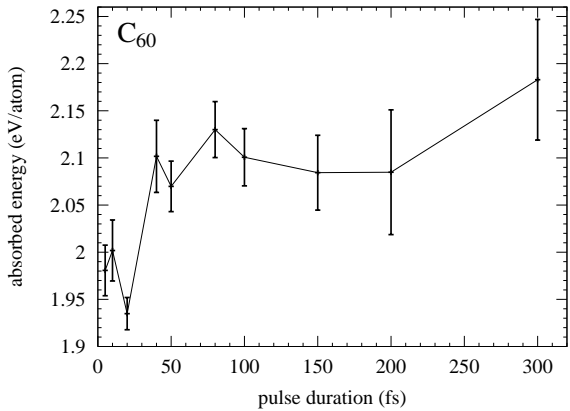}
\vspace*{0.5cm}
\caption{
Nonthermal fragmentation threshold for C$_{60}$ clusters as a function
of laser pulse duration. In the range from $\tau=5$~fs to
$\tau=300$~fs, no clear dependence of the threshold on pulse duration
is observed. A damage threshold of $t_d=2.1\pm0.2$~eV can be
attributed to the entire range of pulse durations.}
\label{fig:threshold}\end{figure}

We have analyzed the nonthermal fragmentation of C$_{60}$ clusters for
a large range of pulse durations and absorbed energies. We observed
that a certain thermalization of the atomic degrees of freedom starts
to occur at $t_{th} \approx 3$~ps after the pulse maximum.  Therefore we
define the threshold for nonthermal fragmentation as the maximal
absorbed energy for which no fragmentation occurs for $t \leq t_{th}$. In
Fig.~\ref{fig:threshold} we show the dependence of the threshold for
nonthermal fragmentation as a function of the pulse duration. The
fragmentation threshold was found to vary only slightly with pulse
duration. For pulses of $\tau=5$~fs to $\tau=300$~fs duration the threshold
is around $t_{th}=2.1$~eV/atom.  We expect that after redistribution
of the energy deposited in the cluster, thermal fragmentation may take
place later even for energies below the threshold $t_{th}$.

\begin{figure}
\includegraphics[width=0.4\textwidth,bb=90 40 415 720]
{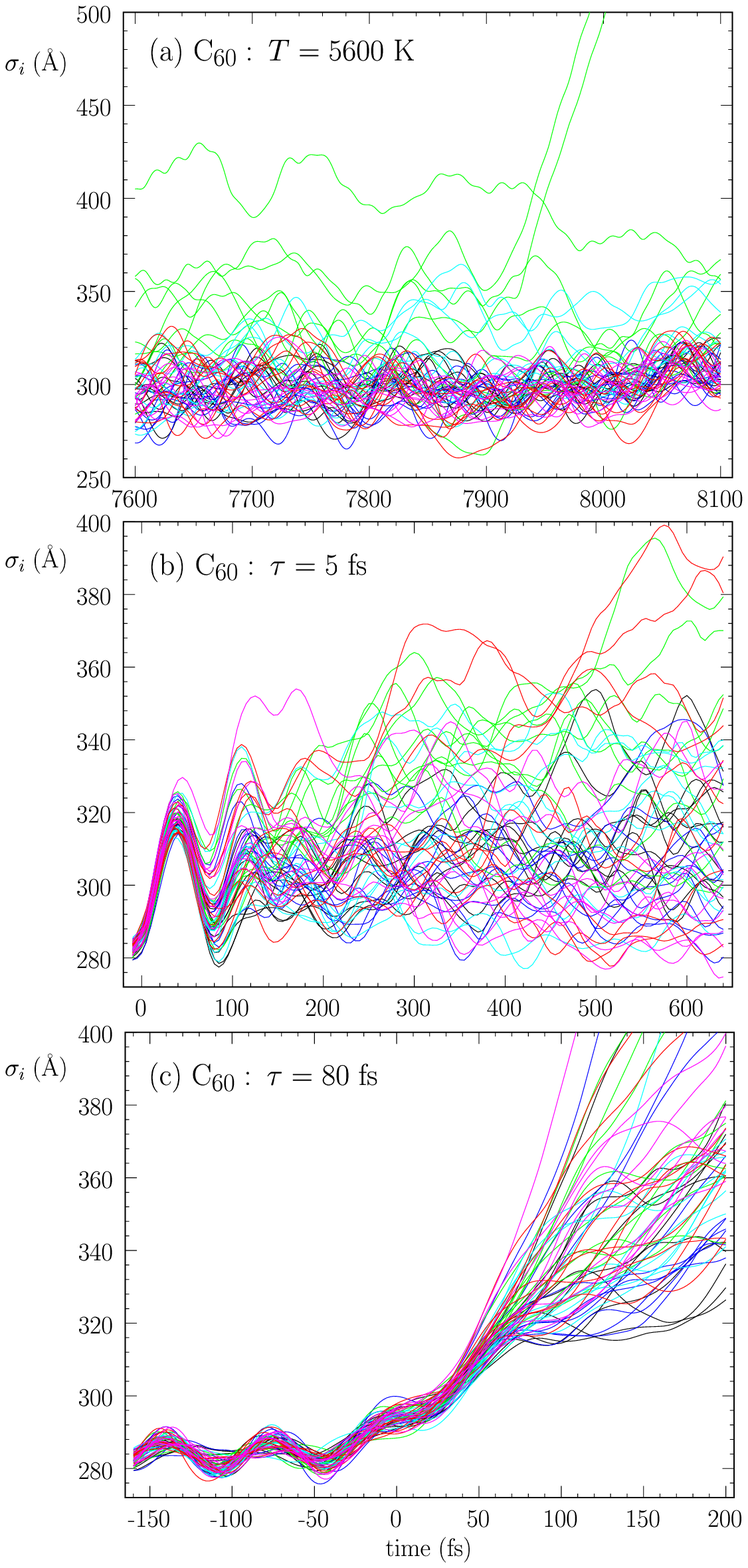}
\caption{ 
Atomic equivalence indices $\sigma_i(t)$ for C$_{60}$ clusters. (a)
Thermal fragmentation at a temperature $T=5600$~K. The abscissa
indicates the absolute time of the trajectory after heating. At
$t=6.8$~ps the first irreparable damage has developed and can be seen
here as the single atomic equivalence indices with values $\sigma_i >
330$~{\AA}\ far above the average $\sigma_i(t)$. They correspond to
carbon chains dangling from the surface of the cluster. At $t=7920$~fs
the emission of a carbon dimer C$_2$ can be observed as two rapidly
increasing atomic equivalence indices.  (b) Isomerization in response
to a $\tau=5$~fs laser pulse. The time evolution of the atomic
equivalence indices corresponds to the same trajectory as the
snapshots in Fig.~\ref{fig:5fs}. Immediately after the laser pulse
maximum at the time $t=0$ the cluster expands strongly. At $t=50$~fs
the coherent motion of the cluster atoms dissolves into a broad
spectrum of individual movements. Atomic equivalence indices with
values above $\sigma_i \simeq 320$~{\AA}\ correspond to chains of
atoms that are dangling at the surface of the original cage structure.
(c) Fragmentation in response to a $\tau=80$~fs laser pulse. The
$\sigma_i$ evolution corresponds to the cluster in
Fig.~\ref{fig:80fs}. An energy $E_0 = 3.5$~eV/atom was absorbed from
the pulse.  The time $t=0$~fs corresponds to the peak of the laser
pulse.  }
\label{fig:sigma}\end{figure}

As mentioned above, it has been found experimentally that carbon
dimers as products of the thermal fragmentation of C$_{60}$ dominate
the fragmentation spectra \cite{smalley:87}. Calculations have been
reported~\cite{scuseria:94} which show that heating of C$_{60}$ to a
temperature of $T = 5600$~K leads to a fragmentation of C$_{60}$ by
the emission of a dimer C$_2$. In order to test the validity of the
nonequilibrium fragmentation of C$_{60}$, we thermalized C$_{60}$
clusters to a temperature $T = 5600$~K. We also find that the thermal
fragmentation process is a dimer emission on a time scale of several
picoseconds. In order to visualize the differences between thermal and
nonthermal fragmentation, we now analyze the different trajectories
with the help of the atomic equivalence
indices~\cite{sugano:89,koutecky:97}.  These quantities, which are
instructive for the characterization of the vibrational excitation and
structural changes in C$_{60}$, are defined by
\begin{equation}
\sigma_i(t) = \sum_j|{\bf r}_i(t)-{\bf r}_j(t)| \,,
\end{equation}
where ${\bf r}_i(t)$ is the position of atom $i$. Thus, the set of
coordinates of the C$_{60}$ cluster yields at each time $t$ a set of
60 atomic equivalence indices $\sigma_i(t)$. For every atom, $\sigma_i$ contains
the structural information of its surroundings. Degeneracies of the
$\sigma_i(t)$ are related to the symmetry of the molecule. In the case of
an undamaged C$_{60}$ molecule the high symmetry of the structure
leads to a time development of all atomic equivalence indices in a
narrow bundle. The atomic equivalence indices corresponding to a
trajectory of a C$_{60}$ molecule at a temperature of $T=5600$~K are
shown in Fig.~\ref{fig:sigma}~(a). The amplitudes of the oscillations
of the single atomic equivalence indices $\sigma_i$ are approximately
$A=22$~{\AA}, compared to $A\approx 7$~{{\AA}} in the case of a C$_{60}$ cluster at
$T=300$~K.  Single atomic equivalence indices with values of $330$~{\AA}\
or more correspond to a dangling chain of carbon atoms that have torn
themselves free from the closed cage of the molecule. These atomic
equivalence indices do not show harmonic oscillations as most of the
$\sigma_i$ of the molecule. In the right side of Fig.~\ref{fig:sigma}~(a),
the emission of a carbon dimer C$_2$ can be seen, corresponding to the
two approximately parallel $\sigma_i$ lines with rapidly increasing
magnitude. Note that the emission process is purely thermal; the
number of electrons thermally excited above the Fermi level was always
below $1\:\%$.

In order to illustrate the dramatic qualitative differences between
thermal and nonthermal fragmentation we show in
Fig.~\ref{fig:sigma}~(b) atomic equivalence indices $\sigma_i(t)$
corresponding to the ultrafast isomerization of C$_{60}$ upon
excitation with a laser pulse of $\tau=5$~fs (see
Fig.~\ref{fig:5fs}). Beginning at the peak of the very short laser
pulse at $t = 0$~fs, the cluster expands strongly. However, this
coherent motion is quickly resolved into an incoherent oscillation of
the $\sigma_i$. Then, $\sigma_i$ lines with large values emerge at $t = 120$~fs
and they do not oscillate harmonically. This indicates damage of the
closed cage structure. The range of the $\sigma_i$ values widens for
subsequent times in accordance with the formation of a protruding
carbon chain that was already mentioned in the description of
Fig.~\ref{fig:5fs}.

Fig.~\ref{fig:sigma}~(c) shows the time development of atomic
equivalence indices $\sigma_i(t)$ during the fragmentation of C$_{60}$ (see
Fig.~\ref{fig:80fs}). During the comparatively long laser pulse of $\tau
= 80$~fs duration, the cluster exhibits a coherent breathing mode, but
shortly after the pulse maximum at $t = 0$~fs a strong expansion of
the cluster is observed. In this case the energy deposited in the
cluster was so high that the disintegration sets in already at $t=
70$~fs, recognizable by the rapidly increasing distances between the
atomic equivalence indices $\sigma_i(t)$. The fact that only a small part
of the $\sigma_i$ lines stay relatively close to each other corresponds to
the fact that a very open linear chain structure has formed in
Fig.~\ref{fig:80fs}.

Summarizing, we have shown that the excitation of C$_{60}$ with
femtosecond pulses gives rise to a nonthermal response which is
qualitatively different from the well known thermal emission of
dimers. Since we consider only electronic excitations below the
continuum states, a comparison of our with existing experimental
results based on detection of ionic fragments is difficult. However,
experimental fragment mass
spectra\cite{hertel1,hertel2,campbell,hunsche} show clearly the
existence of fragments of different sizes and a particularly large
peak for C$^+$, which would confirm our simulations. In order to check
that single ionization does not affect the main predictions of this
work we have performed calculations on C$^+_{60}$ clusters and we
obtain essentially the same nonthermal fragmentation thresholds.

This work has been supported the Deutsche Forschungsgemeinschaft
through SFB 450, and DGESIC (Grant PB98-0345). One of us (M.~E.~G.) 
acknowledges an Iberdrola Visiting Professorship at the University of
Valladolid.

\end{multicols}
\end{document}